\documentclass[dvips]{article}
\usepackage{supertabular,lscape,epsfig}
\usepackage{amssymb}
\usepackage{amsmath}

\DeclareSymbolFont{ppa}{OT1}{ppl}{m}{it}
\DeclareMathSymbol{\vv}{\mathalpha}{ppa}{'166}

\begin{document}
\newcommand{\via}{{\it via},\,}

\newcommand{\dd}{\,{\rm d}}
\newcommand{\ie}{{\it i.e.},\,}
\newcommand{\etal}{{\it et al.\ }}
\newcommand{\eg}{{\it e.g.},\,}
\newcommand{\cf}{{\it cf.\ }}
\newcommand{\vs}{{\it vs.\ }}
\newcommand{\zdot}{\makebox[0pt][l]{.}}
\newcommand{\up}[1]{\ifmmode^{\rm #1}\else$^{\rm #1}$\fi}
\newcommand{\dn}[1]{\ifmmode_{\rm #1}\else$_{\rm #1}$\fi}
\newcommand{\upd}{\up{d}}
\newcommand{\uph}{\up{h}}
\newcommand{\upm}{\up{m}}
\newcommand{\ups}{\up{s}}
\newcommand{\arcd}{\ifmmode^{\circ}\else$^{\circ}$\fi}
\newcommand{\arcm}{\ifmmode{'}\else$'$\fi}
\newcommand{\arcs}{\ifmmode{''}\else$''$\fi}
\newcommand{\MS}{{\rm M}\ifmmode_{\odot}\else$_{\odot}$\fi}
\newcommand{\RS}{{\rm R}\ifmmode_{\odot}\else$_{\odot}$\fi}
\newcommand{\LS}{{\rm L}\ifmmode_{\odot}\else$_{\odot}$\fi}

\newcommand{\Abstract}[2]{{\footnotesize\begin{center}ABSTRACT\end{center}
\vspace{1mm}\par#1\par
\noindent
{~}{\it #2}}}

\newcommand{\TabCap}[2]{\begin{center}\parbox[t]{#1}{\begin{center}
  \small {\spaceskip 2pt plus 1pt minus 1pt T a b l e}
  \refstepcounter{table}\thetable \\[2mm]
  \footnotesize #2 \end{center}}\end{center}}

\newcommand{\TableSep}[2]{\begin{table}[p]\vspace{#1}
\TabCap{#2}\end{table}}

\newcommand{\FigCap}[1]{\footnotesize\par\noindent Fig.\  %
  \refstepcounter{figure}\thefigure. #1\par}

\newcommand{\TableFont}{\footnotesize}
\newcommand{\TableFontIt}{\ttit}
\newcommand{\SetTableFont}[1]{\renewcommand{\TableFont}{#1}}

\newcommand{\MakeTable}[4]{\begin{table}[htb]\TabCap{#2}{#3}
  \begin{center} \TableFont \begin{tabular}{#1} #4 
  \end{tabular}\end{center}\end{table}}

\newcommand{\MakeTableSep}[4]{\begin{table}[p]\TabCap{#2}{#3}
  \begin{center} \TableFont \begin{tabular}{#1} #4 
  \end{tabular}\end{center}\end{table}}

\newenvironment{references}%
{
\footnotesize \frenchspacing
\renewcommand{\thesection}{}
\renewcommand{\in}{{\rm in }}
\renewcommand{\AA}{Astron.\ Astrophys.}
\newcommand{\AAS}{Astron.~Astrophys.~Suppl.~Ser.}
\newcommand{\ApJ}{Astrophys.\ J.}
\newcommand{\ApJS}{Astrophys.\ J.~Suppl.~Ser.}
\newcommand{\ApJL}{Astrophys.\ J.~Letters}
\newcommand{\AJ}{Astron.\ J.}
\newcommand{\IBVS}{IBVS}
\newcommand{\PASP}{P.A.S.P.}
\newcommand{\Acta}{Acta Astron.}
\newcommand{\MNRAS}{MNRAS}
\renewcommand{\and}{{\rm and }}
\section{{\rm REFERENCES}}
\sloppy \hyphenpenalty10000
\begin{list}{}{\leftmargin1cm\listparindent-1cm
\itemindent\listparindent\parsep0pt\itemsep0pt}}%
{\end{list}\vspace{2mm}}

\def\TYLDA{~}
\newlength{\DW}
\settowidth{\DW}{0}
\newcommand{\dw}{\hspace{\DW}}

\newcommand{\refitem}[5]{\item[]{#1} #2%
\def\REFARG{#3}\ifx\REFARG\TYLDA\else, {\it#3}\fi
\def\REFARG{#4}\ifx\REFARG\TYLDA\else, {\bf#4}\fi
\def\REFARG{#5}\ifx\REFARG\TYLDA\else, {#5}\fi.}

\newcommand{\Section}[1]{\section{#1}}
\newcommand{\Subsection}[1]{\subsection{#1}}
\newcommand{\Acknow}[1]{\par\vspace{5mm}{\bf Acknowledgements.} #1}
\pagestyle{myheadings}

\newfont{\bb}{ptmbi8t at 12pt}
\newcommand{\xrule}{\rule{0pt}{2.5ex}}
\newcommand{\xxrule}{\rule[-1.8ex]{0pt}{4.5ex}}
\newcommand{\uprule}{\rule{0pt}{2.5ex}}
\newcommand{\douprule}{\rule[-2ex]{0pt}{4.5ex}}
\newcommand{\dorule}{\rule[-2ex]{0pt}{2ex}}
\begin{center}
{\Large\bf The Optical Gravitational Lensing Experiment.
Real Time Data Analysis Systems in the OGLE-III Survey.\footnote{Based
on observations obtained with the 1.3~m
Warsaw telescope at the Las Campanas  Observatory of the Carnegie
Institution of Washington.}}

\vskip1cm
{\bf A.~~U~d~a~l~s~k~i}
\vskip3mm
{Warsaw University Observatory, Al.~Ujazdowskie~4, 00-478~Warszawa,
Poland\\
e-mail: udalski@astrouw.edu.pl}
\end{center}

\Abstract{We describe three real time data analysis systems implemented
during the third phase of the OGLE survey (OGLE-III). The EWS system is
designed to detect on-going microlensing events. The EEWS system monitors
the microlensing phenomena for anomalies from the single mass
microlensing. The NOOS system detects transient objects in the OGLE-III
fields (SNe, microlensing events, variable stars) that normally are below
the detection threshold. Information on objects detected by each of these
systems is distributed to the astronomical community for follow up
observations. Also a short description of the OGLE-III hardware and
photometric data pipeline is presented.}{}

\Section{Introduction}
In June 2001, after the second major hardware upgrade, the Optical
Gravitational Lensing Experiment entered its third phase, OGLE-III, and
with the data flow of about 3.5~TB/year became a ``terabyte'' survey.  With
more than 200 millions of stars observed regularly once every 1--3 nights
it became crucial to perform the photometric reductions in real time to
avoid saturation by the huge incoming data stream. While the initial
reductions of the OGLE-III images were performed in real time from the very
beginning, the full on-line photometric data pipeline was implemented one
year later for the Galactic bulge fields when sufficient number of good
quality frames were collected for each field. Gradually, all the remaining
OGLE-III fields were included for on-line photometric reductions and
presently all images are reduced within several minutes after the frame has
been collected.

From the point of view of the scientific output of a sky survey it is also
very crucial to implement systems that can analyze the photometry of the
observed fields in real time. Many objects undergo large brightness
variations in different time scales and detection of such cases,
potentially very interesting, in early phases can be very important for
good sampling or follow up observations.

One of such systems, the Early Warning System (EWS) for detection of
on-going microlensing events was designed and installed in 1994 during the
OGLE-I phase, as the first system of such type in any microlensing
survey. Later similar capabilities were implemented by the MACHO group
(Alcock \etal 1997), in very limited range by the EROS group (Afonso
\etal 2003) and since 2000 by the MOA Collaboration (Bond \etal 2001).
The EWS system operated successfully during OGLE-II providing about 40--80
microlensing detections per year.

The modified EWS system was reinstalled for OGLE-III in May 2002 and
operated successfully in 2002 and 2003 Galactic bulge observing seasons.
Two other real time data analysis systems were also developed and
implemented in the past months: EEWS -- a system of detection of
anomalies in microlensing events and NOOS -- a system of detection of
transient objects in the OGLE-III fields. In this paper we present
description of all these real time systems and shortly describe the data
acquisition system and data photometry pipeline of the OGLE-III survey.

\vskip 8pt
\Section{OGLE-III Data Acquisition System}
\vskip 8pt
Replacement of the single chip CCD camera with a new ``second''
generation instrument was the main feature of the hardware upgrade of
OGLE-III. The new CCD camera is an eight chip ${8192\times8192}$ pixel
mosaic. Each chip of the mosaic is a SITe ST-002a CCD detector with
${2048\times4096}$ pixels of 15~$\mu$m size. This corresponds to the
0.26 arcsec/pixel scale in the focus of the OGLE telescope, and the full
field of view of the mosaic is ${35\times35}$~arcmins.

The chips are mounted in two rows on the molybdenum mounting plate in large
cryogenic dewar (ND10) manufactured by IR Laboratories Inc. (Tucson, AZ,
USA). They are cooled and held at the temperature of ${-95}$~C. Part of
electronics, namely preamplifiers and bias and clock signal filters are
also mounted inside the dewar, very close to the detectors. Only one CCD
amplifier is used for reading of each chip. Therefore the following
electronics consist of eight identical parallel channels. The liquid
nitrogen (LN2) hold time of the dewar (capacity of 6 liters) is about 24
hours.

The signals are directed to and from the dewar electronics {\it via} two
hermetic 100 pin connectors. Two small aluminum boxes mounted very close to
these connectors hold small printed circuit boards with clock
electronics. These boxes are connected with short high quality cables with
the larger ``gold'' box mounted to the instrument base, containing the main
electronics boards. The electronics of each channel consists of two parts:
DC chain, providing all bias and clock voltages and signal chain: dual
slope integrator with LTC1608 ADC converter for signal digitalization. Each
channel (output voltages, ADC converter, gain etc.) is programmable by a
microcontroller based on digital signal processor (DSP) TMS320C50. Also
clocking pattern can be programmed by this microcontroller. The electronics
is adjusted in this manner that in the default mode all CCD chips have
similar gain value. The readout noise of chips depends on the chip and is
in the range of 6--9~e$^-$. The well depth reaches from 60\,000~e$^-$ to
80\,000~e$^-$. Table~1 lists the gain and readout noise values for each
chip. The default clocking pattern allows reading of the mosaic in about 98
seconds.
\MakeTable{ccc}{12.5cm}{Gain and readout noise of CCD detectors}
{\hline
\uprule CHIP   &    GAIN    &  READOUT NOISE\\
\dorule        & [e$^-$/ADU]&    [e$^-$]\\
\hline
\uprule
1      &    1.31    &      6.7\\
2      &    1.36    &      6.3\\
3      &    1.37    &      6.2\\
4      &    1.34    &      9.2\\
5      &    1.32    &      5.9\\
6      &    1.33    &      6.7\\
7      &    1.33    &      7.1\\
8      &    1.34    &      8.7\\
\hline}

The microcontroller is identical with that used in OGLE-II (Udalski,
Kubiak, Szyma{\'n}ski 1997) and includes additional DSP board for reading
the output of ADC converter of each channel during the next pixel reading
time. The data are then transmitted {\it via} the TAXI line and PC-PCI
board to the PC computer located outside the telescope level. This computer
can operate locally, but in the normal observing mode it immediately sends
the data received from the microcontroller over the GB Ethernet network to
the main data acquisition computer located in the control room a few tens
meters outside the telescope building. The data acquisition computer reads
the incoming stream of data, assigns the coming pixels to appropriate chips
and stores the data on the disk in eight independent FITS files of about 17
MB each. They share the same image number but the extension corresponds to
the chip number. The total size of each image (eight files) is about 137
MB.
 
The remaining hardware components of the system, namely the guider, filter
wheel and shutter are identical as during OGLE-II (Udalski, Kubiak,
Szyma{\'n}ski 1997). Only minor modifications to the shutter and filter
wheel were necessary to avoid vignetting.
\vspace*{-10pt}
\Section{OGLE-III Photometric Data Reduction Pipeline}
\vspace*{-5pt}
The automatic software designed for on-the-fly flat-fielding of the
collected frames continuously monitors output from the data acquisition
software waiting for the incoming images and begins the operation when a
current image is completely read. Each chip in the image is treated
separately. The images are de-biased and flat-fielded by a procedure based
on the standard IRAF\footnote{IRAF is distributed by National Optical
Observatories, which is operated by the Association of Universities for
Research in Astronomy, Inc., under cooperative agreement with National
Science Foundation.} routines from {\sc CCDPROC} package. Appropriate
bias and flat-field images for each chip from the mosaic are prepared in
advance by independent code, also based on {\sc CCDPROC} package. To save
hard disk space, flat-fielded and raw images are finally compressed with
the RICE algorithm. Once every few days they are dumped to a HP Ultrium
tape (about 200~GB of raw data per tape) for storage and transportation to
the headquarters.

When the current image is preprocessed, the status of its field is verified
and if the image is supposed to be reduced in real time then it is
processed by the main OGLE-III photometric data pipeline. The photometry
software applies the image subtraction method. It is based on the
Wo{\'z}niak's (2000) implementation of the Difference Image Analysis, DIA,
technique (Alard and Lupton 1998, Alard 2000) with many software
modifications for better performance and stability.

Similarly to the initial pre-processing, images from each chip are
processed independently. Depending on the stellar density of the field an
image is divided into thirty two ${512\times512}$, eight ${1024\times1024}$ or two
${2048\times2048}$ pixel subframes. In the first step of reductions the
shift between the frame and the reference image (see below) is calculated
and the subframes of the current image are extracted and interpolated to
the same grid of coordinates as the subframes of the reference image. Then
the transformation between the current subframe and reference image
subframe is derived and the difference image subframe (current minus
reference) is created.

In the next step the difference image of the current subframe is searched
for objects that brightened or faded. Next, the positions of these
detections are cross-correlated with the positions of stars detected in the
reference image and two files containing the known variable stars and
``new'' variable stars (those whose $X,Y$ coordinates do not correlate with
known stars) in the current difference image are created. Finally, the
photometry of all objects identified earlier in the reference image at the
position of their centroids is derived.  Photometric measurement consists
of the measured PSF flux of the star in the difference image (Wo{\'z}niak,
2000) added to the PSF flux of this object in the reference image,
converted then to the magnitude scale.

When all subframes of the current image are reduced, the individual
files for each subframe are combined into three global files of a given
frame containing current photometry of all reference image stars,
detected variable objects and ``new'' objects in the current frame. One
has to remember that many of the ``new'' objects are artifacts (traces of
non-perfectly removed cosmic-rays, non-perfectly subtracted bright stars
etc.).

In the final step the photometric databases are updated. The databases are
created and updated with the software similar to that used in previous
stages of the OGLE project (Szyma{\'n}ski and Udalski 1993) with some minor
modifications due to somewhat different data input format. Separate
databases are created for each chip in a given field, thus the complete set
for a field consists of eight separate databases.  Moreover, two kinds of
databases are created in each case -- the large database contains
photometry of all stars detected in the reference image while the second
one, much smaller, includes only ``new'' objects that are detected in the
subsequent images. While the number of objects is fixed in the former case
it gradually increases in the latter. The databases ensure fast and user
friendly access to the complete photometry of all OGLE-III objects.

Before a field can be included in the list of the fields whose photometric
reductions are performed in real time, a reference image and appropriate
files with magnitudes of stellar objects detected in this image must be
constructed. The reference image is constructed by stacking and averaging
several single good quality and seeing images. The first image on the list
of components of the reference image defines the pixel grid of the
reference image. Depending on the stellar density in the field the images
are divided into 32, 8 or 2 subframes. This division is then preserved
during the reduction procedures. Each subframe is treated independently.
Before the reference image of a subframe is constructed the shifts of all
images on the list of components of the reference image relative to the
first image are calculated and appropriate subframes are extracted. Each
subframe is cleaned for cosmic ray hits and all known defects of the
detector are masked.

Next, precise transformations between the subframe of the first image and
the remaining subframes are derived, the subframe pixels are interpolated
to the same pixel grid and the scale factor is derived. If the scale is
considerably different than the scale of the first image,
\eg due to poor atmospheric transparency, the subframe is removed. In
the next step all subframes from the list are averaged after multiplying
by the scale factor and correcting for difference in the background
level. Finally, the PSF photometry of the subframe of the reference
image is derived using the {\sc DoPHOT} photometry program (Schechter,
Saha and Mateo 1993). The flux values obtained with PSF photometry are
converted to the scale of the DIA software based on the linear
transformation obtained from several tens of the brightest stars. The
procedure is repeated for all subframes of a given field.

The reference image has much deeper magnitude range than the individual
images. Therefore the PSF photometry obtained from this image is much
more accurate and the list of detected stars is much more complete. In
the Galactic bulge fields the reductions are performed on thirty two
${512\times512}$ subframes while in the Magellanic Cloud fields on eight
${1024\times1024}$ or two ${2048\times2048}$ pixel subframes depending
on the number of stars in the field. Images of each chip in a given
field are divided in the same manner.
\vspace*{-10pt}
\Section{Early Warning System (EWS)}
\vspace*{-5pt}
The EWS system of real time detection of microlensing events in progress
was implemented in 1994 during the first phase of the OGLE project (Udalski
\etal 1994) as the first system of such type in microlensing surveys.
Later -- in 1998 -- it was adapted for the OGLE-II data pipeline and
operated successfully to the end of OGLE-II (end of 2000). In short, the
EWS system compared the current brightness of a star with its mean
brightness in the reference observing season and registered all objects
that brightened more than a threshold for five consecutive times. The
threshold depended on the magnitude of star and was derived as three times
of the typical magnitude {\it rms} of non-variable stars at a given
brightness. Then, marked objects were further analyzed and promising
candidates for microlensing events (both single mass or binary
microlensing) were alerted.

The EWS system in OGLE-III underwent significant modifications compared to
the previous OGLE-I and OGLE-II implementation. New photometric data
pipeline of OGLE-III, based on the DIA photometry, required a new algorithm
for detection of objects varying in brightness. Also, because of large
number of monitored stars, of the order of hundreds of millions, the
filtering algorithm had to be efficient enough to filter out unavoidable
artifacts. The large number of possible artifacts could make smooth
operation of the system difficult or practically impossible.

Similarly to old versions, the OGLE-III EWS system operates on stars that
are detected in the reference images. In this manner all detected
microlensing events have well established baseline brightness what makes
their further analysis simpler and more reliable. Because the stars are
detected on the reference frames that are the average of several co-added
good seeing individual frames, the stellar detection threshold is much
lower than in the individual images making the stellar lists much more
complete.

Difference image resulting from subtraction of the current and reference
images serves as an indicator of stellar variability in the field. Output
files from the photometric data pipeline that contain variable objects
detected in the current difference image which were cross-identified with
the entries in the list of stars from the reference image, are used by
EWS. The appropriate flags are set in the so called variability index
created for each field and each chip. The variability index is updated at
the same time when the current frame is added to the photometric database
and contains easy to search information on the photometric behavior of a
given star in the field. The entries for each star in the index contain the
total number of points when the star was detected as bright or faint in the
difference images and the current number of consecutive measurements when
the star was bright (faint). Because detection of objects in the difference
images is seeing dependent, the number of consecutive measurements allows
for one missing detection before it is set back to zero. To avoid large
number of very low variability bright objects an additional constraint on
variable stars is imposed, namely the brightening or fading of a star has
to be larger than 0.06~mag. This limit effectively works only for the
brightest stars. For fainter stars the detection in the difference image
provides a natural detection threshold. It is somewhat lower than the
detection limit used in the OGLE-I and OGLE-II version of the EWS system.

The variability index provides additional field for masking variable stars
and artifacts. Because the fields observed during OGLE-III are seasonal,
after each observing season all objects that were detected during that
season in the difference images more than once, either brighter or fainter
than in the reference image, are flagged. In this manner most of the
periodic or long term variable stars and also artifacts (artificial
variable objects often close to saturated stars etc.) can be removed from
the sample monitored for microlensing events. Separate flagging of the
consecutive observing seasons makes the index useful not only for the
detection of current variability but also for analyzing the past
variability.

To select microlensing event candidates the EWS software scans the
variability indices after each observing night. When the number of
consecutive observations of a star brighter than the threshold is larger
than a preassigned number, currently ${N=2}$, and the star was non-variable
in the reference season(s), the object is marked for further analysis. The
light curves of these stars are then inspected visually. The vast majority
of marked objects are artifacts resulting, for instance, from contamination
of the measurements by nearby bright and severely saturated stars, CCD
detector defects, non-perfectly removed cosmic ray hits etc. Also
observations at very poor seeing conditions can contaminate photometry
(poor image subtraction) leading to artificial detections in some
cases. New artifacts can also be masked using additional flag in the
variability index.

Selected microlensing event candidates showing the light curves that
resemble the light curves of microlensing events are further verified in
images. It is checked whether the star indeed increased its
brightness. Candidates that passed this step are considered as possible
microlensing events. The automatic software prepares the photometry
dataset, finding charts, plots of the light curve, appropriate files for
the ftp archive, WWW page for each event and the EWS announcement. When all
candidates from a night are processed the announcement is e-mailed to the
EWS mailing list subscribers.

The performance of the EWS system strongly depends on the quality of
filtering. The OGLE-III EWS system became operational in the beginning of
May 2002 for the 2002 Galactic bulge season. Unfortunately, the number of
observations collected for the Galactic bulge in the previous (2001) season
was relatively small (of the order of 10--20), because less than half of
the observing period was covered. Therefore during the 2002 season the
filtering of variable stars and artifacts was not tight enough and large
number of false objects were usually selected for visual inspection after
each night. Due to large number intensive masking of the marked variable
stars and artifacts was necessary, practically after each night, to ensure
reasonable operation of the system. A side effect of such a mode of
operation were possible errors in identification of real microlensing
effects. The best example is the spectacular high magnification
microlensing event MOA 2002-BLG-33 alerted by the MOA group on 2002, June
18 (Abe \etal 2003). The same event (catalog name BLG205.1 121022) was
triggered by the EWS system at very early phases of lensing on 2002,
May 12. The candidate was overlooked among hundreds of artifacts, then
masked and in this manner remained undetected by the EWS system. Another
drawback of so early operation of the system was relatively large
contamination by variable stars of microlensing candidates announced in the
2002 season, reaching 11\%. For instance, short baselines did not allow to
exclude cataclysmic variables -- a few evident dwarf novae were alerted as
possible microlensing candidates: 2002-BLG-077, 2002-BLG-090, 2002-BLG-119,
2002-BLG-129. Also many microlensing events were announced relatively late
-- near or after maximum -- and therefore they were not suitable for follow
up observations by other microlensing groups. Nevertheless, the overall
operation of the system in 2002 was successful -- the total number of
detected microlensing events (about 350) was comparable with the total
number of events discovered during the MACHO or OGLE-II surveys.

The operation of the EWS system was suspended at the end of the 2002
Galactic bulge season in November 2002. It was resumed at the beginning of
May 2003 for the 2003 Galactic bulge season. Due to much longer time span
available for filtering out variable stars and artifacts (2001 and 2002
seasons) operation of EWS during the 2003 bulge season was very smooth as
the filtering was much more effective than in 2002 season. Typically the
number of objects for visual inspection after each night was about 100--300
for about 100 million photometric measurements. With so small number of
triggered objects flagging the artifacts was not necessary and therefore
the chance of missing an event was considerably smaller than during the
2002 season. Contamination of alerted candidates by variable stars was also
smaller than in the 2002 season, although a few dwarf novae again mimicked
binary microlensing events: 2003-BLG-034, 2003-BLG-271. The total number of
alerted microlensing event candidates in the 2003 season reached about 460.

One should remember that even with perfect filtering the EWS system may
miss some microlensing events. The system requires now three detections of
the brightening of a star to mark it as a candidate. The typical sampling
of the Galactic bulge fields in the middle of the season is on average once
per 1.5--2 nights and less frequent at the beginning or the end of the
season. Therefore very short time scale events might be missed or
discovered well after maximum, in particular when the regular sampling of
the fields is affected by longer periods of bad weather, passage of the
Moon close to the bulge fields (observations are typically suspended for
2--3 nights) or other reasons.

The EWS system will operate in similar mode with some minor adjustments
during the following Galactic bulge seasons. A small number of new fields
will be added at the beginning of the 2004 season while some fields with no
or small number of microlensing detections will not be monitored
anymore. Also starting from the beginning of 2004 all OGLE-III Magellanic
Cloud fields will be monitored by the EWS system.  Altogether the EWS
system monitors more than 200 million stars: 33 and 170 millions in the
Magellanic Clouds and the Galactic bulge, respectively.

Information on the current status of the EWS system microlensing
candidates can be found from the following addresses:
\vspace*{-7pt}
\begin{itemize}
\itemsep=-5pt
\item OGLE main page:
 
\centerline{{\it http://ogle.astrouw.edu.pl/}}
\centerline{{\it http://bulge.princeton.edu/\~{}ogle}}
\item EWS system main page:
\vskip3pt
\centerline{{\it http://ogle.astrouw.edu.pl/ogle3/ews/ews.html}}
\item EWS system ftp address:
\vskip3pt
\centerline{{\it ftp://ftp.astrouw.edu.pl/ogle/ogle3/ews}}
\end{itemize}
One can subscribe to the EWS mailing list from the main EWS page.

\Section{Early Early Warning System (EEWS)}
Observations of a disturbance in the standard light curve of a single mass
microlensing event can provide additional and very important information on
the lensing system. While some of the disturbances occur in long time
scales (\eg a parallax effect) and can be easily detected with regular
survey mode sampling of the observed fields (provided the photometric
accuracy is good enough) some other may have time scale of hours and can be
easily overlooked or poorly sampled with regular observing pattern of the
OGLE-III survey. Of particular interest are disturbances induced by
planetary companions of lensing stars or first caustic crossing in binary
microlensing. In the latter case the moment of the first caustic crossing
is highly unpredictable contrary to the second caustic crossing which time
can be estimated from earlier observations. The passage through the
caustics lasts typically a few hours so it is essential to observe the
event every few minutes. Good coverage of both caustic crossings is very
important for unambiguous modeling of a binary event. In the case of
disturbances caused by planetary companions of the lensing star its time
scale can also be of the order of hours and again frequent observations of
the event are crucial for correct interpretation of the event. The best
example here is the microlensing event 2002-BLG-055 where a single data
point deviation from single mass microlensing was observed on the falling
part of the light curve. Modeling of the light curve suggested possible
planetary signal (Jaroszy{\'n}ski and Paczy{\'n}ski 2002) but due to scarce
coverage of the light curve at the moment of the deviation other
interpretations of the event cannot be ruled out.

The Early Early Warning System (EEWS) is another OGLE-III system of data
analysis working in real time aiming at the detection of deviations from
the regular single mass microlensing light curve profile. The main goal of
the system is to provide the OGLE observer fast information on the current
behavior of already discovered microlensing events and enabling the
observer fast response, \ie changing the regular survey mode to follow-up
mode of frequent observations of a particular event.

The EEWS system works in the following way. When a new frame is collected
and then flat-fielded by the automatic procedure (Section~3), the EEWS
daemon selects the microlensing events from this field using the list of
already known microlensing events in the OGLE-III fields. The list
includes all well observed events with light curves of good enough quality
from all OGLE-III seasons. The list is regularly updated when new events
are discovered by the EWS system. Presently (after 2002 and 2003 observing
seasons) the list comprises about 470 objects.

For each lens from the field independent reduction procedures are
performed on the subframe of the current frame where the lens is located
in. A separate dedicated for the EEWS system CPU is used for photometric
reduction. The reduction procedure is identical with that of the
standard OGLE-III photometric pipeline, but due to small size of the
subframe the reduction takes just a few seconds. In the next observing
seasons this step will not be necessary, because after a recent upgrade
the current CPU power available in OGLE-III is large enough that
standard pipeline reduction of a frame is finished within a few minutes
after it has been collected.

When the current magnitude of a lens is derived it is compared with the
magnitude predicted by the single mass model light curve fitted to the
previous observations. If the difference is larger than three times the
error of the observation or a preassigned threshold (0.15~mag), whichever
larger, the EEWS system prepares the appropriate plots and data files and
e-mails a "RED ALERT" to the observer. After visual inspection of the
current frame, difference frame, the light curve of the event etc.\ the
observer undertakes further action if the alert is promising: it switches
the observing program from the survey mode to the follow-up mode. The
deviating lens is observed more frequently with the sampling appropriate to
changes of its brightness. Typically, the current light curve and
photometry is available to the observer after 1--4 minutes after
observation.

The EEWS system was implemented during the 2003 Galactic bulge season at
the end of May 2003. Soon after that the first important alert was made,
namely the EEWS system alerted fast brightening of the microlensing event
2003-BLG-170. The brightening turned out to be a crossing of the first
caustic in binary microlensing. Due to the "RED ALERT" from the EEWS the
caustic crossing, impossible to predict in advance, was very well
sampled. Another examples of spectacular events alerted by the EEWS
system include 2003-BLG-238 (deviation due to finite size of the source
star) and 2003-BLG-267 microlenses (again crossing of the first caustic in
binary microlensing). Fig.~1 presents the light curve of OGLE 2003-BLG-267.
\begin{figure}[htb]
\vglue-4mm
\centerline{\includegraphics[width=11cm, bb=20 50 510 540]{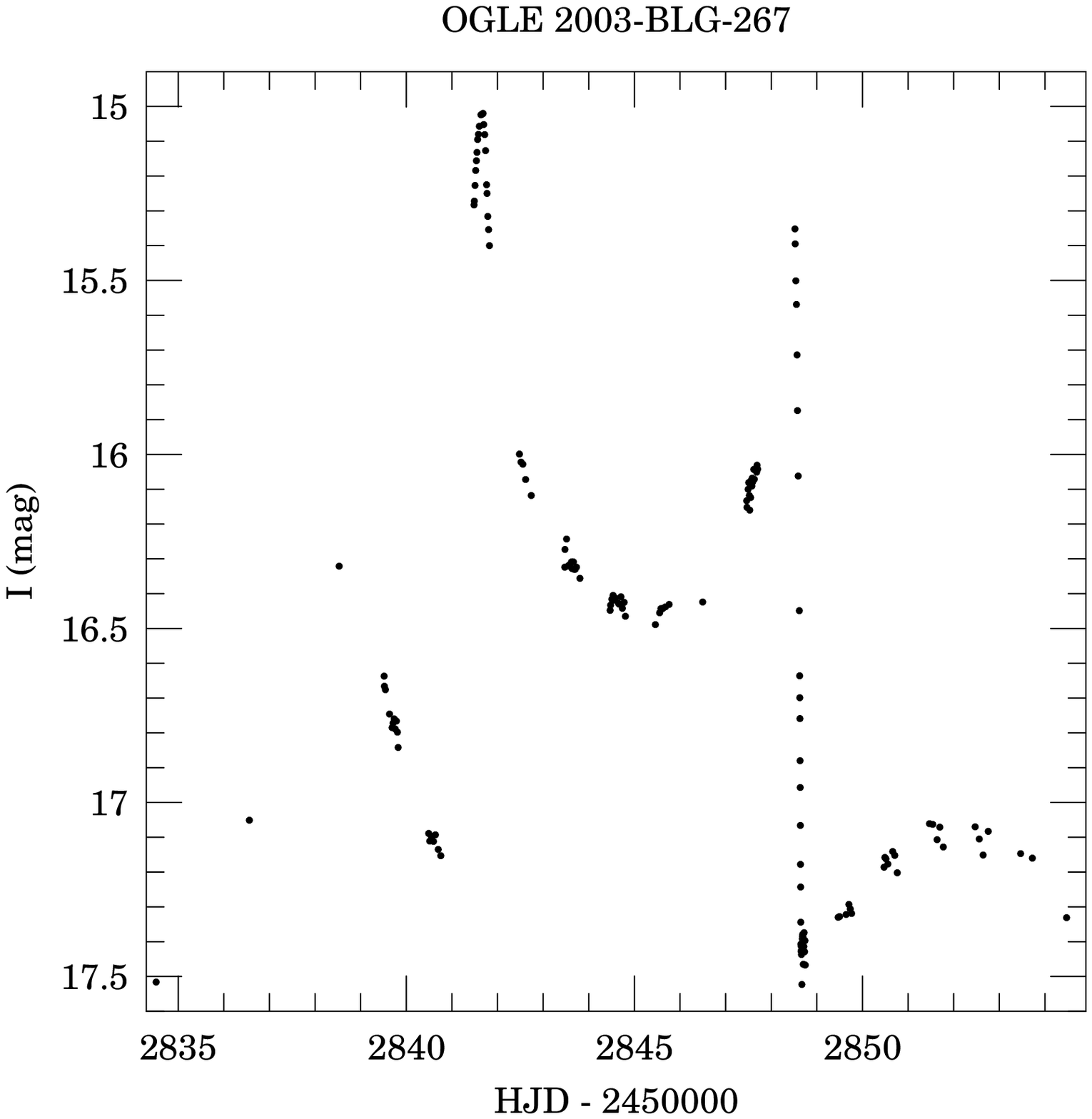}}
\FigCap{Light curve of OGLE 2003-BLG-267 binary microlensing event. The
event was alerted by the EEWS system on HJD=2839.5 enabling good sampling
of crucial phases of the event.}
\end{figure}

Operation of the EEWS system during the 2003 Galactic bulge season
proved that it works effectively and is very important tool for
increasing flexibility of the OGLE-III survey to the events requiring
fast time response. One has, however, to remember that the sensitivity
of the system is limited at early phases of microlensing events, \ie
before the maximum of light. At these stages the microlensing fit to the
observations is usually poorly constrained, in particular for fainter
stars or when the gaps between observations are larger due to, for
instance, bad weather. Therefore, it is not always possible to recognize
and correctly interpret a new deviating observation. Very often the old
observations combined with a new deviating one provide a microlensing
fit of similar quality to the old one (with different parameters),
especially when the amplitude of the deviation is small. In this manner
it is easy to overlook an interesting deviation. The situation is much
better after the maximum of microlensing event. Then, the fit to the
large part of normally behaving single point microlensing is so well
constrained that even a small magnitude deviation can be easily
recognized. For example, were the EEWS system implemented in 2002
observing season, the 2002-BLG-055 lens deviation from single mass
microlensing would certainly be detected and alerted and more
observations of this possible planetary disturbance would be collected
at crucial moments.

The EEWS system will be used by OGLE-III during the next observing
seasons. Similarly to the 2003 season we plan to distribute EEWS alerts
{\it via} our EWS network informing the microlensing follow-up groups about
very interesting events requiring instantaneous observing response.

\Section{New Objects in the OGLE Sky (NOOS)}
The EWS and EEWS systems operate on stars previously detected in reference
images of the OGLE-III fields. Because the reference image is a co-added
image of several best individual images its range is deeper than that of
regular frames. However, if objects that are fainter than the detection
threshold of our reference images increase their brightness they can be
temporarily seen in the OGLE frames. Such transient objects can potentially
be very interesting. Therefore another real time data analysis software was
designed and implemented in OGLE-III: the New Objects in the OGLE Sky
(NOOS) system.

The main goal of the NOOS system is the detection of transient objects in
the OGLE-III fields. After each observing night, when the OGLE databases
are updated, the NOOS scans the databases of ``new'' objects looking for
entries that have more than selected number of detections (currently
${N=2}$). Such objects are marked for further analysis. Next, the automatic
software prepares the finding charts for each candidate -- a part of the
reference image of a given field with position of the analyzed transient
marked. These finding charts are visually inspected for objects whose
positions do not coincide with nearby bright (often saturated) stars,
defects on the detector etc. For promising candidates the most recent
frames are then inspected to verify whether the detection is real. The
remaining objects are masked so they do not contaminate further NOOS
operation.

If the candidate is accepted then its photometry is derived by re-running
the reduction procedures (identical to the OGLE standard photometric
pipeline) and measuring the flux of the transient at its ``bright''
position. Fifteen images prior the detection are typically measured. Next,
the light curve plots, finding charts, as well as the WWW page and ftp
archive files for the object are prepared. An alert notifying subscribers
about the discovery is also prepared and distributed {\it via} the OGLE NOOS
alert mailing list.

The NOOS system was implemented in the middle of November 2003. The
Magellanic Cloud fields were the first targets, altogether about 53 square
degrees around both Clouds. During the first month of operation more than
ten new transient objects were detected in these fields. Additional ten
were found in the earlier data -- collected from the beginning of the
2003/2004 season (September 2003). The typical detection threshold of a
transient is ${I\approx19.8}$~mag. The typical lag of detection is
about 6--10 days after a transient becomes brighter than the detection
threshold. Good sampling of the observed fields (once every 2--4 nights)
ensures well covered light curves of the transient objects.
\begin{figure}[htb]
\hglue-6mm{\includegraphics[width=13.3cm, bb=100 85 520 560]{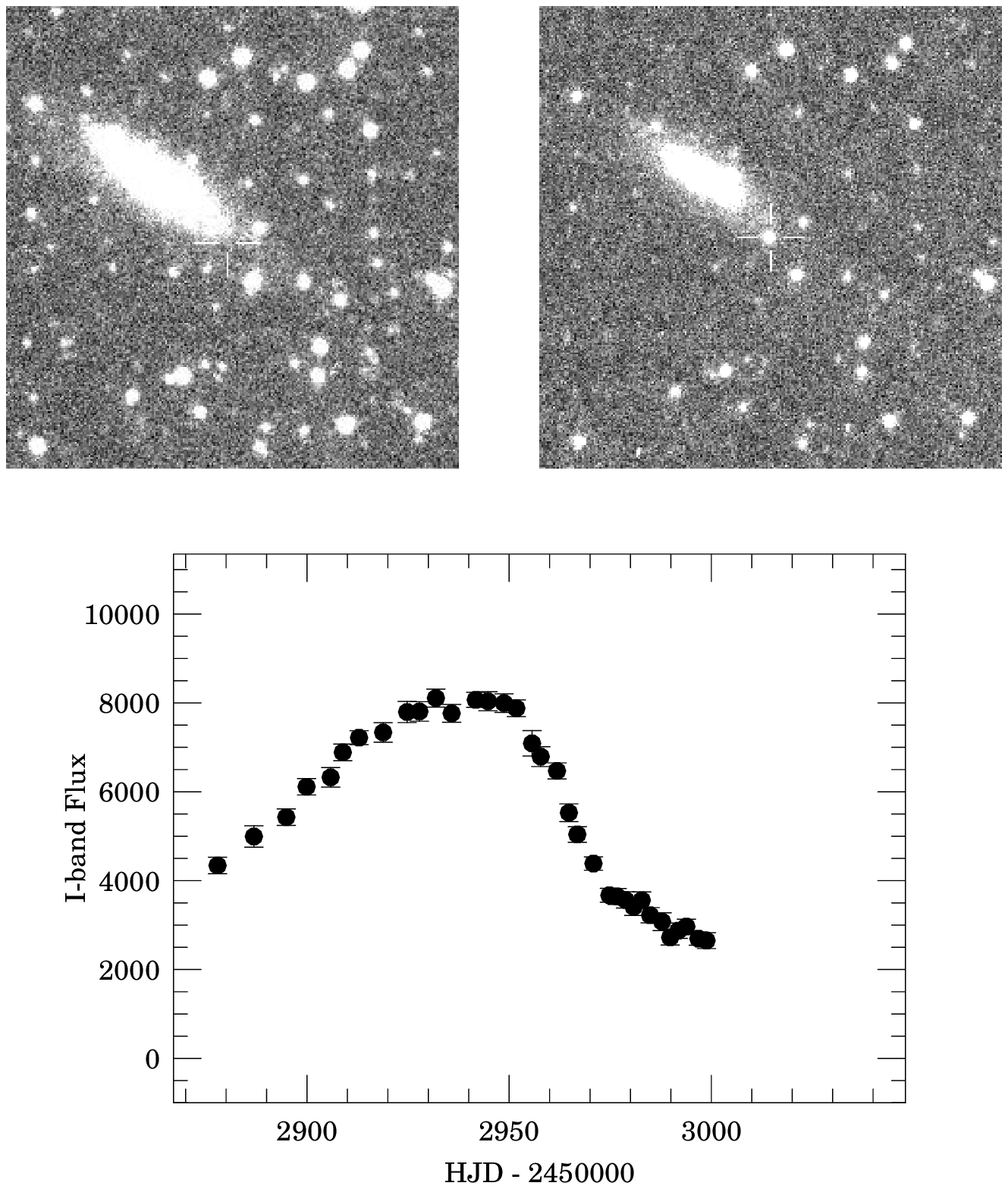}}
\FigCap{Field and light curve of the SN OGLE 2003-NOOS-005.}
\end{figure}
\vskip 20pt
\begin{figure}[htb]
\hglue-6mm{\includegraphics[width=13.3cm, bb=100 85 520 560]{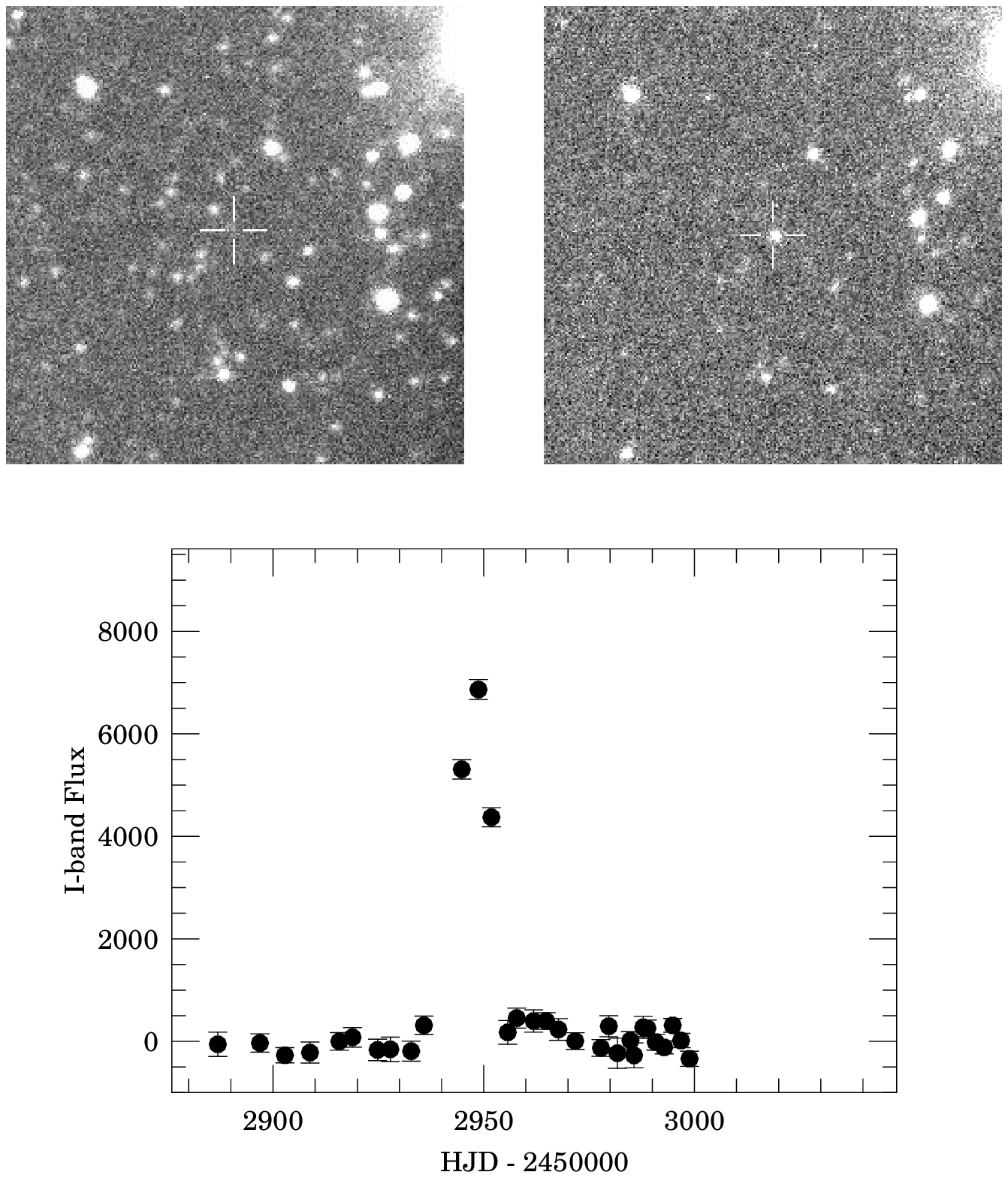}}
\FigCap{Field and light curve of the transient OGLE 2003-NOOS-011.}
\end{figure}

The sample of the transients detected so far indicates that the full
variety of objects may belong to this class: SNe, high magnification
microlensing events of faint stars, cataclysmic variables or other variable
stars of large amplitude. Sometimes the light curve may provide a clue to
classification. For instance, several transient objects detected by the
NOOS system are certainly SNe, usually detected by NOOS before maximum of
light (\eg 2003-NOOS-005, 2003-NOOS-006, 2003-NOOS-012, 2003-NOOS-016,
2003-NOOS-020). However, in many cases follow up observations are needed,
in particular spectroscopy. Therefore, we encourage observers to follow up
the OGLE--NOOS variables. Anything that suddenly brightens is potentially
very interesting and worth observing. As an example of transient objects
the light curve of a SN -- 2003-NOOS-005 -- is shown in Fig.~2, and a
transient -- 2003-NOOS-011 -- in Fig.~3.

The NOOS system will also cover the Galactic bulge fields starting from the
2004 Galactic bulge season. It is expected that the vast majority of
transients from this region will be high magnification microlensing events
of faint Galactic bulge stars.

Information about the transients detected by the NOOS system can be found
from the following addresses:
\begin{itemize}
\item OGLE main page:
\begin{center}
{\it http://ogle.astrouw.edu.pl/}  \\
{\it http://bulge.princeton.edu/\~{}ogle} 
\end{center}
\item NOOS system main page: 
\begin{center}
{\it http://ogle.astrouw.edu.pl/ogle3/ews/NOOS/noos.html}
\end{center}
\item NOOS system ftp address: 
\begin{center}
{\it ftp://ftp.astrouw.edu.pl/ogle/ogle3/ews/NOOS/}
\end{center}
\end{itemize}
Astronomers interested in receiving e-mail notification on new discoveries
may subscribe to the NOOS mailing list from the main page of the NOOS
system.

\Acknow{I would like to thank Drs.\ M.\ Szyma{\'n}ski, B.\ Paczy{\'n}ski 
and M.\ Kubiak for their help at all stages of the OGLE-III phase of OGLE
survey. I am also very grateful to Dr.\ A.\ Szentgyorgy for his valuable
help in obtaining electronic parts for the OGLE-III mosaic camera. The
paper was partly supported by the Polish KBN grant 2P03D02124 to A.\
Udalski. Partial support to the OGLE project was provided with the NSF
grant AST-0204908 and NASA grant NAG5-12212 to B.~Paczy{\'n}ski.
I acknowledge support from the grant ``Subsydium Profesorskie'' of the
Foundation for Polish Science.}

\end{document}